\documentclass{ws-mpla}

\begin{document}

\markboth{F. Paradis, H. Kr\"oger, X.Q. Luo, K.J.M. Moriarty}
{Monte Carlo Hamiltonian of Lattice Gauge Theory}

\catchline{}{}{}{}{}
%
%%%%%%%%%%%%%%%%%%%%%%%%%%%%%%%%%%%%%%%%%%%%%%%%%%%%%%%%%%%%%%%%%%%%

\title{MONTE CARLO HAMILTONIAN OF LATTICE GAUGE THEORY}

\author{F. PARADIS}

\address{D\'epartment de Physique, Universit\'e Laval, \\
Qu\'ebec, Qu\'ebec, G1K 7P4, Canada \\
email: fparadis@phy.ulaval.ca}

\author{H. KR\"OGER\footnote{Corresponding author}}

\address{D\'epartment de Physique,  Universit\'e Laval, \\
Qu\'ebec, Qu\'ebec, G1K 7P4, Canada \\
email: hkroger@phy.ulaval.ca}

\author{X.Q. LUO}

\address{Department of Physics, Zhongshan University, \\
Guangzhou 510275, People's Republic of China \\
email: stslxq@zsu.edu.cn }

\author{K.J.M. MORIARTY}

\address{Institute for Computational Studies, \\
Department of Mathematics, Statistics and Computational Science \\
Dalhousie University, Halifax, Nova Scotia B3H 3J5, Canada \\
email: kevin.moriarty@cs.dal.ca }

\maketitle

\pub{Received (Day Month Year)}{Revised (Day Month Year)}

\begin{abstract}
We discuss how the concept of the Monte Carlo Hamiltonian
can be applied to lattice gauge theories.

\end{abstract}

\ccode{PACS Nos.: 11.15.-q, 03.70.+k}

\section{Bargman states}

In standard lattice field theory it is customary to compute
vacuum expectation values of operators, like
\begin{equation}
\label{eq:VacVacCorrFct}
<\Omega,t=+\infty| A(x,t) B(x',t') |\Omega,t=-\infty> ~ ,
\end{equation}
which denotes the amplitude that the system is in its ground
state $|\Omega>$ at $t=-\infty$, is acted upon by some local
field operators at some intermediate time and returns to the
ground state at $t=+\infty$. In lattice calculations, using a
lattice of finite extension in space and time, one can extract
from such correlators  information about particle masses in
the ground state and in the lowest lying excited states.
In order to extract information about particle properties in
higher excited states one may consider the following amplitudes
\begin{equation}
\label{eq:BargmanTransAmp}
<\Psi_{fi},t=T|\Psi_{in},t=0> ~ ,
\end{equation}
where $|\Psi>$ denotes a Bargman state. A Bargman state on a
space-time lattice at a given time slice is defined by
assigning a particular value of a local field variable to each
lattice node. In the case of the real scalar bosonic field
$\Phi(x,t)$, for instance, in a given time-slice $t$ each lattice
node $x_{k}$ is associated with some real number $\Phi_{k}$
(taking values between $+\infty$ to $-\infty$). This may be
visualized as a regular array in a plane with perpendicular
spikes (of different length) poking out of the plane on either
side. The ground state can be considered as a linear superposition
of many Bargman states. On the other hand, a single Bargman state
can be considered as a linear superposition of the ground state
plus of many excited states. Thus, by computing matrix elements
like in Eq.(\ref{eq:BargmanTransAmp}), one expects to get
information on the sector of excited states of the system.
The reason why we have chosen Bargman states in particular,
is that those are localized states, which is very convenient
in local quantum field theory.

\section{How to compute transition amplitudes by Monte Carlo?}
The idea is to express the quantum transition amplitude in terms
of a path integral and to evaluate such path integral numerically, using Monte Carlo with importance sampling. However, such Monte Carlo method allows only to compute the ratio of transition amplitudes, like
\begin{equation}
\label{eq:RatioTransAmp}
R = \frac{ <\Psi_{fi},t=T|\hat{O}(t=T_{int})|\Psi_{in},t=0> }
{ <\Psi_{fi},t=T|\Psi_{in},t=0> } ~ ,
\end{equation}
but not a single transition amplitude. A solution of this problem
has been proposed for quantum mechanical systems\cite{Jirari99},
and also for the scalar model in field theory\cite{Huang02}. The idea is to split the action occuring in the path integral into a free, i.e. kinetic term plus an interacting, i.e., potential term. As an example let us consider a 1-d quantum system of a massive particle in some potential
\begin{eqnarray}
\label{eq:SplitActionQM}
S[x] &=& \int_{0}^{T} dt \frac{1}{2} m \dot{x}^{2} - V(x)
= \int_{0}^{T} dt \frac{1}{2} m \dot{x}^{2} - \int_{0}^{T} dt V(x)
\equiv S^{kin}[x] + S^{pot}[x] ~ .
\end{eqnarray}
Then the transition matrix element in imaginary time can be expressed (we use $\hbar=1$ throughout)
\begin{eqnarray}
\label{eq:SplitPathIntQM}
< z| exp[-H T]|y> &=& < z| exp[-H^{kin} T]|y>
\nonumber \\
&\times&
\frac{
\left.
\int [dx] ~ \exp[ - S^{pot}[x] ] ~ \exp[ -S^{kin}[x] ] \right|^{z,T}_{y,0} }
{ \left.
\int [dx] ~ \exp[ -S^{kin}[x] ] \right|^{z,T}_{y,0} } .
\end{eqnarray}
Here $\exp[ - S^{pot}]$ is treated as an observable and
$\exp[ -S^{kin}]$ represents a weight factor.
The ratio of path integrals can be computed by standard Monte Carlo methods with importance sampling.
The matrix element $< z| exp[-H^{kin} T]|y>$, describing the free propagator (in imaginary time) is known analytically,
\begin{equation}
\label{eq:FreeTransAmplQM}
< z| exp[-H^{kin} T]|y> =
\sqrt{ \frac{ m }{ 2 \pi T } } ~
\exp \left[ - \frac{ m}{2 T } (z - y)^{2} \right] .
\end{equation}
The calculation of this transition amplitude proceeds most simply
by using momentum space, which diagonalises $H^{kin}$ given by
\begin{equation}
\label{eq:FreeHamQM}
H^{kin} = \frac{P^{2}}{2m} ~ .
\end{equation}
The connection between position space and momentum space is given by
\begin{equation}
\label{eq:BasisTransfQM}
<x|p> = \frac{1}{\sqrt{2 \pi }} \exp[i p x] ~ .
\end{equation}
Note that this scalar product is solely determined by the algebraic commutator relation of the operators $\hat{X}$ and $\hat{P}$, plus the normalisation of the position and momentum states.

\section{Splitting the action in lattice gauge theory}
Let us consider $QED$ on the lattice without fermions.
The group of gauge symmetry is $U(1)$. We keep in mind how to construct a lattice Hamiltonian (see e.g, Creutz\cite{Creutz}, Rothe\cite{Rothe}) via the transfer matrix, by splitting the lattice action into terms of time-like and space-like plaquettes,
\begin{eqnarray}
\label{eq:SplitGaugeAction}
S[U] &=& \frac{1}{g^{2}}\frac{a}{a_{0}} \sum_{P_{time-like}}
[1 -Re(U_{P})] +
\frac{1}{g^{2}}\frac{a_{0}}{a} \sum_{P_{space-like}}
[1 -Re(U_{P})]
\nonumber \\
&\equiv&  S^{kin}[U] + S^{pot}[U] ~ .
\end{eqnarray}
The transition amplitude can be split up, in analogy to the
example from quantum mechanics, Eq. (\ref{eq:SplitPathIntQM}),
\begin{eqnarray}
\label{eq:SplitPathIntGauge}
<U_{fi}|\exp[- H T ]| U_{in}> &=&
<U_{fi}|\exp[- H^{kin} T ]| U_{in}>
\nonumber \\
&\times&
\frac{
\left.
\int [dU] ~ \exp[ - S^{pot}[U] ] ~ \exp[ -S^{kin}[U] ] \right|^{U_{fi},T}_{U_{in},0} }
{ \left.
\int [dU] ~ \exp[ -S^{kin}[U]] ]
\right|^{U_{fi},T}_{U_{in},0} } .
\end{eqnarray}
Like in the quantum mechanical example, the ratio of path integrals can be computed using Monte Carlo with importance sampling. The problem is now the computation of the transition amplitude evolving according to the kinetic term.

\section{Transition amplitude of evolution due to kinetic term}
In order to calculate the amplitude
\begin{equation}
\label{eq:FreeTransAmplGauge}
<U_{fi}|\exp[- H^{kin} T ]| U_{in}>
\end{equation}
we make use of the close correspondence between quantum mechanics and lattice gauge theory. There is a lattice Hamiltonian, which is the generator of infinitesimal translations in time, the so-called Kogut-Susskind Hamiltonian. In the case of $U(1)$ gauge theory,
it reads \cite{Rothe}
\begin{equation}
\label{eq:KogutSusskind}
H_{KS} = \frac{g^{2}}{2a} \sum_{<ij>} \hat{l}_{ij}^{2} + \frac{1}{g^{2}a} \sum_{P_{space-like}} [1 - Re(U_{P})] ~ .
\end{equation}
Here $\hat{l}_{ij}$ denotes the operator of the electric field between spatial neighbour lattice sites $i$ and $j$. This operator on the lattice has discrete eigenvalues,
\begin{equation}
\label{eq:ElectField}
\hat{l}_{ij} |\lambda_{ij}> = \lambda_{ij} |\lambda_{ij}> ~ , \lambda_{ij} = 0,\pm 1, \pm 2, \dots
\end{equation}
We can express the transition amplitude, Eq.
(\ref{eq:FreeTransAmplGauge}), by
\begin{eqnarray}
\label{eq:FreeTransAmplGauge2}
&& <U_{fi}|\exp[- H^{kin} T ]| U_{in}>
\nonumber \\
\nonumber \\
&=& \sum_{\lambda} <U_{fi}|\lambda><\lambda|\exp[- H^{kin} T ]
|\lambda><\lambda| U_{in}>
\nonumber \\
&=& \sum_{\lambda,\lambda'} <U_{fi}|\lambda><\lambda|\exp \left[
- \frac{g^{2}T}{2a} \sum_{<ij>} \hat{l}_{ij}^{2} \right]
|\lambda'><\lambda'| U_{in}>
\nonumber \\
&=& \prod_{ij} \left\{ \sum_{\lambda_{ij}=0,\pm1,\pm2,\dots} <U^{fi}_{ij}|\lambda_{ij}>
\exp\left[- \frac{g^{2}T}{2a} \lambda_{ij}^{2} \right]
<\lambda_{ij}|U^{in}_{ij}> \right\} ~ .
\end{eqnarray}
Due to the locality of the Bargman-link state and the structure of the kinetic part of the Kogut-Susskind Hamilton operator, this amplitude factorizes in separate amplitudes for each link.
In order to evalute such individual link amplitudes one needs to compute
the scalar product which defines the transition from the link basis to the electric field basis,
\begin{equation}
\label{eq:BasisTransfGauge}
<U_{ij}|\lambda_{ij}> ~ .
\end{equation}

\subsection{Scalar product $<U|\lambda>$}
Let us compute the scalar product given by Eq.
(\ref{eq:BasisTransfGauge}). For simplicity of notation, we drop
the subscript $ij$ in this section. Like in quantum mechanics,
this scalar product is determined solely by the algebraic
commutator relation between the operators $\hat{U}$ and $\hat{l}$,
plus the normalisation of the link and electric field states. Let
us recall the properties of the link basis \cite{Creutz},
\begin{eqnarray}
\label{eq:PropLinkBasis}
&& \hat{U}|U> = U|U>
\nonumber \\
&& <U'|U> = \delta(U'-U)
\nonumber \\
&& 1 = \int ~ dU ~ |U><U| ~ ,
\end{eqnarray}
where $\int ~ dU$ denotes a group integral and $\delta(U'-U)$ denotes a
$\delta$ function (distribution) over group elements.
Likewise, the electric field basis obeys
\begin{eqnarray}
\label{eq:PropElectrBasis}
&& \hat{l}|\lambda> = \lambda|\lambda>
\nonumber \\
&& <\lambda'|\lambda> = \delta_{\lambda',\lambda}
\nonumber \\
&& 1 = \sum_{\lambda=0,\pm1,\pm2,\dots} |\lambda><\lambda| ~ .
\end{eqnarray}
The algebraic commutator relation of $\hat{l}$ and $\hat{U}$ is given by\cite{Creutz,Rothe}
\begin{equation}
\label{eq:Commutator}
[\hat{l},\hat{U}]= - \hat{U} ~ .
\end{equation}
First, we want to compute the matrix element $<U'|\hat{l}|U>$.
Using Eq. (\ref{eq:Commutator}), we have
\begin{equation}
<U'|[\hat{l},\hat{U}]|U> = <U'|\hat{l}\hat{U}|U> - <U'|\hat{U}\hat{l}|U>
= [U - U']<U'|\hat{l}|U> ~ ,
\end{equation}
and on the other hand
\begin{equation}
<U'|[\hat{l},\hat{U}]|U> = <U'|- \hat{U}|U> = -U <U'|U> = -U
\delta(U' - U) ~ .
\end{equation}
Thus we find
\begin{equation}
\label{eq:MatElemL}
<U'|\hat{l}|U> = -U \frac{\delta(U'-U)}{U' - U}
= (-) U \frac{\partial}{\partial U'} \delta(U'-U) ~ ,
\end{equation}
where we used the following result from distribution theory
\begin{equation}
\label{eq:DerivDeltaDistr}
\frac{\delta(x'-x)}{x' - x}
= - \frac{\partial}{\partial x'} \delta(x'-x) ~ .
\end{equation}
Second, we want to compute the matrix element $<\lambda|U>$. Using
Eq. (\ref{eq:Commutator}), we obtain
\begin{equation}
\label{eq:LUStep1}
<\lambda|[\hat{l},\hat{U}]|U> = <\lambda|- \hat{U}|U>
= - U <\lambda|U> ~ ,
\end{equation}
and also
\begin{eqnarray}
\label{eq:LUStep2}
&&<\lambda|[\hat{l},\hat{U}]|U> =
\lambda U <\lambda|U> - <\lambda|\hat{U}\hat{l}|U>
\nonumber \\
&& = \lambda U <\lambda|U>
- \int dV <\lambda|\hat{U}|V><V|\hat{l}|U>
\nonumber \\
&& = \lambda U <\lambda|U>
+ \int dV ~ U V <\lambda|V> \frac{\partial}{\partial V}\delta(V - U)
\nonumber \\
&& = \lambda U <\lambda|U> - \frac{\partial}{\partial V}
\left. \left[ UV <\lambda|U> \right] \right|_{V=U}
\nonumber \\
&& = \lambda U <\lambda|U>
- \left[ U <\lambda|U> + U^{2} \frac{\partial}{\partial U}<\lambda|U> \right] ~ .
\end{eqnarray}
Combining Eqs. (\ref{eq:LUStep1},\ref{eq:LUStep2}) yields the
differential equation
\begin{equation}
\label{eq:LUDiffEq}
U \frac{\partial}{\partial U}<\lambda|U> = \lambda <\lambda|U> ~ .
\end{equation}
The solution has the form
\begin{equation}
\label{eq:LUGenSol}
<\lambda|U> = B_{\lambda} (U)^{\lambda} ~ ,
\end{equation}
where $B_{\lambda}$ is some constant. This constant is determined
from the normalisation condition of the electric field states, Eq.
(\ref{eq:PropElectrBasis})
\begin{eqnarray}
\label{eq:NormCond}
&&\delta_{\lambda',\lambda} = <\lambda'|\lambda>
= \int dU <\lambda'|U><U|\lambda>
\nonumber \\
&&= B_{\lambda'} (B_{\lambda})^{*} \int dU ~ (U)^{\lambda' - \lambda}
= B_{\lambda'} (B_{\lambda})^{*} \frac{1}{2\pi} \int d\alpha
\exp[i \alpha(\lambda' - \lambda)]
\nonumber \\
&&= B_{\lambda'} (B_{\lambda})^{*} \delta_{\lambda',\lambda} ~ ,
\end{eqnarray}
Here we have used the orthogonality of periodic functions
\begin{equation}
\label{eq:Orthogonal}
\int_{-\pi}^{+\pi} d \xi e^{in\xi} e^{-in' \xi}
= 2 \pi \delta_{n,n'} ~ .
\end{equation}
We have parametrized a link $U = \exp(i \alpha)$, with the group
integral measure (Haar measure) $dU = {d \alpha}/(2 \pi)$. From
Eq. (\ref{eq:NormCond}) we conclude
\begin{equation}
\label{eq:BValue}
B_{\lambda}=1 ~ .
\end{equation}
From Eqs. (\ref{eq:LUGenSol},\ref{eq:BValue}) we conclude
\begin{equation}
\label{eq:LUSol}
<\lambda|U> = (U)^{\lambda} ~ .
\end{equation}

\subsection{Transition amplitude for a single link}
Now we obtain the transition amplitude for a single link, denoting
$U_{fi}=\exp(i \alpha_{fi})$ and $U_{in}=\exp(i \alpha_{in})$
\begin{eqnarray}
\label{eq:FreeTransAmplGaugeSingleLink}
&& <U_{fi}|\exp[- H^{kin} T ]| U_{in}>
\nonumber \\
\nonumber \\
&=& \sum_{\lambda=0,\pm1,\pm2,\dots} <U^{fi}|\lambda>
\exp[- \frac{g^{2}T}{2a} \lambda^{2} ]
<\lambda|U^{in}>
\nonumber \\
&=& \sum_{\lambda=0,\pm1,\pm2,\dots}
\exp[- \frac{g^{2}T}{2a} \lambda^{2} ]
\exp[i \lambda (\alpha_{in} - \alpha_{fi})]
\nonumber \\
&=& \sum_{n=0,\pm1,\pm2,\dots}
\exp[- \frac{g^{2}T}{2a} n^{2} ]
\cos[n(\alpha_{in} - \alpha_{fi})] ~ .
\end{eqnarray}
\\

\noindent Let us do a consistency check of this result. Let us
consider the time evolution during a time interval $T=a_{0}$
(lattice spacing in time) and then take the limit $a_{0}\to 0$.

We define
\begin{eqnarray}
\label{eq:Definition}
&&\Delta \alpha = \alpha_{fi} - \alpha_{in}
\nonumber \\
&&A^{2} = \frac{g^{2} a_{0}}{2 a}
\nonumber \\
&&x_{n} = A n, ~~~ n=0,\pm 1,\pm2,\dots, ~~~ \Delta x = A ~ .
\end{eqnarray}
Then the transition amplitude, Eq.
(\ref{eq:FreeTransAmplGaugeSingleLink}) becomes
\begin{eqnarray}
\label{eq:TransAmplGaugeSingleStep} <U_{fi}|\exp[- H^{kin} a_{0}
]| U_{in}> &=& \frac{1}{\Delta x} \sum_{n=0,\pm1,\pm2,\dots}
\Delta x ~ \exp[- x_{n}^{2} ] \cos[x_{n} \Delta \alpha /\Delta x]
\nonumber \\
&\sim_{\Delta x \to 0}& \frac{1}{A}
\int_{-\infty}^{+\infty} dx ~ \exp[-x^{2}]
\cos(x \Delta \alpha/A)
\nonumber \\
&=& \frac{\sqrt{\pi}}{A} \exp[- \frac{(\Delta \alpha)^{2}}{4 A^{2}} ]
\nonumber \\
&=& \sqrt{\frac{2\pi a}{g^{2} a_{0}}}
\exp[- \frac{a}{a_{0}g^{2}} \frac{(\Delta \alpha)^{2}}{2}]
\nonumber \\
&\sim_{a_{0} \to 0}& 2\pi \delta(\alpha_{fi} - \alpha_{in})
= \delta(U_{fi} - U_{in}) ~ ,
\end{eqnarray}
which reproduces the correct behavior in the limit when $a_{0}$ goes to zero. Moreover, in the second to last line one may replace
$(\alpha_{fi} - \alpha_{in})^{2}/2$ by
$1 - \cos[\alpha_{fi} - \alpha_{in}]$. This is justified
in distribution theory when $a_{0}$ goes to zero. Then one obtains the expression of the path integral of link variables for a single time step.

It is interesting to see an application of this approach to
non-abelian lattice gauge theory and dynamical quarks. Hamiltonian
lattice QCD at finite chemical
potential\cite{Gregory:1999pm,Luo:2000xi,Luo:2004mc} might be a
more natural solution to the infamous complex action problem.

\vspace*{6pt}

\noindent {\bf Acknowledgment.} F. Paradis, H. Kr\"oger and K.J.M. Moriarty
have been supported by NSERC Canada. X.Q. Luo has been supported
by Key Project of NSF China (10235040), and Project of the Chinese
Academy of Sciences (KJCX2-SW-N10) and Key Project of National
Ministry of Eduction (105135) and Guangdong Ministry of Education.

\end{document}